# Biomechanical Lower Limb Model to Predict Patellar Position Alteration after Medial Open Wedge High Tibial Osteotomy

Elaheh Elyasi[a*], Antoine Perrier[a,b,c], Mathieu Bailet[c], Yohan Payan[a]

[a] *Univ. Grenoble Alpes, CNRS, TIMC-IMAG, 38000 Grenoble, France;*

[b] *Service de chirurgie osseuse et traumatologique, centre de référence des infections ostéo-articulaires complexes, groupe hospitalier Diaconesses–Croix Saint-Simon, 125, rue d'Avron, 75020 Paris, France;*

[c] *TwInsight, 38000 Grenoble, France.*

**\*Corresponding author. Email:** elyasi.elaheh@gmail.com

Laboratoire TIMC-IMAG, Faculté de Médecine, Pavillon Taillefer, 38706 La Tronche Cedex, France

Phone: +33 7 63 73 08 70

**Authors' email addresses:**

Antoine Perrier: antoine.perrier@twinsight-medical.com

Yohan Payan: yohan.payan@univ-grenoble-alpes.fr

Mathieu Bailet: mathieu.bailet@twinsight-medical.com

Keywords: knee; patellofemoral; finite element; multibody; High tibial osteotomy

**Abstract**

Medial open-wedge high tibial osteotomy is a surgical treatment for patients with a varus deformity and early-stage medial knee osteoarthritis. Observations suggest that this surgery can negatively affect the patellofemoral joint and change the patellofemoral kinematics. However, what causes these effects and how the correction angle can change the surgery's impact on the patellofemoral joint has not been investigated before. The objective of this study was to develop a biomechanical model that can predict the surgery's impact on the patellar position and find the correlation between the opening angles and the patellar position after the surgery. A combined finite element and multibody model of the lower limb was developed. The model's capabilities for predicting the patellofemoral kinematics were evaluated by performing a passive deep flexion simulation of the native knee and comparing the outcomes with magnetic resonance images of the study subject at various flexion angles. The model at a fixed knee flexion angle was then used to simulate the high tibial osteotomy surgery virtually. The results showed a correlation between the wedge opening angles and the patellar position in various degrees of freedom. These results indicate that larger wedge openings result in increased values of patellar distalization, lateral patellar shift, patellar rotation, and patellar internal tilt. The developed model in this study can be used in future studies to monitor the stress distribution on the patellar cartilage and connecting tissues to investigate their relationship with observations of pain and cartilage injury due to post-operative altered patellar kinematics.

## 1. INTRODUCTION

Medial open-wedge high tibial osteotomy (OWHTO) is a treatment for patients with a varus deformity who are suffering from early-stage medial knee osteoarthritis (Agarwala et al., 2016; Duivenvoorden et al., 2014; Loia et al., 2016). OWHTO surgery involves a wedge osteotomy on the medial side of the proximal tibia starting just above the level of the tibial tubercle (Hernigou et al., 1987; Noyes et al., 2006). The surgery aims to correct limb alignment and has been shown to provide satisfactory long-term outcomes (Aglietti et al., 2003; Coventry et al., 1993). It successfully relieves the patient's pain and prevents secondary osteoarthritis leading to a total knee arthroplasty (Sun et al., 2017).

Conversely, there are observations indicating that the patellofemoral pressure increases after OWHTO (Javidan et al., 2013; Stoffel et al., 2007), and patellofemoral cartilage injuries progress in 41-45 % of patients in second-look arthroscopy (Goshima et al., 2017; Kim et al., 2017). Additionally, it has been shown that performing large deformity corrections (wedge opening angles greater than 9-10°) may cause degeneration of patellofemoral cartilage (Otakara et al., 2019; Tanaka et al., 2019). Therefore, patellofemoral cartilage injuries seem to be correlated with the size of the opened wedge. The cause of this effect appears to be related to the impact of wedge opening on the patellofemoral kinematics.

The systematic review performed by Elyasi et al., 2021, has shown that after OWHTO, a significant decrease in the patellar height, patellar lateral tilt, and patellar shift is reported in various clinical studies (Bito et al., 2010a; J. C. H. Fan, 2012; Lee et al., 2016; Longino et al., 2013; Park et al., 2017a; Song et al., 2012). Meanwhile, using a method based on Magnetic Resonance Imaging (MRI) to assess the three-dimensional patellofemoral kinematics has shown that the alteration of the kinematics after OWHTO cannot be assessed by means of conventional radiology (standing radiographs) and using the patellar position assessment methods designed on that basis (such as

Insall-Salvati ratio and Blackburne-Peal ratio) (d'Entremont et al., 2014; Elyasi et al., 2021). Therefore, it has not yet been possible to correlate the deterioration of patellofemoral cartilage status with the patellar position due to the limitations of the investigation methods and, more importantly, due to the fact that a quantitative relationship between the size of the opened wedge and the exact patellar position in its various degrees of freedom is missing in the literature.

Computational models of the knee joint have shown their ability to predict patellofemoral kinematics once validated against cadaveric or in vivo experiments (Ali et al., 2020, 2016; Erdemir et al., 2012; Müller et al., 2020). Consequently, this study aimed to develop a combined Finite Element (FE) and multibody model of the lower limb capable of investigating the impact of OWHTO surgery on the patellar position. The patellar position which was described in the patellofemoral joint coordinate system, was used to investigate its correlation with the opening angle. This development was a necessary step to numerically assess the stress distribution on patellar cartilage in the future and to find its relationship with degeneration of patellofemoral cartilage after OWHTO. Prior to using the model for OWHTO, the accuracy of its predictions was evaluated by performing a passive deep flexion simulation of the native knee. The patellofemoral kinematic outcomes were compared against MRIs of the subject at various knee flexion angles.

## 2. METHODS

### 2.1. Subject information and image acquisition

A healthy subject (male, 40 years old, 94 kg, 1.73 m) volunteered to participate in the experimental data collection as part of a pilot study approved by an ethical committee (MammoBio MAP-VS pilot study N°ID RCB 2012-A00340-43, IRMaGe platform, Univ. Grenoble Alpes). He gave his written informed consent to the experimental procedure. Five sessions of non-weight-bearing MRI using a clinical system (Achieva 3.0T dStream Philips Healthcare) and one session of computed tomography (CT) scan were conducted to acquire the imaging data of the subject's lower limb at different knee flexion angles (Figure 1).

The CT scan was done with the subject oriented in the supine position and in a low energy procedure to limit the risks associated with exposure. The scanner collected 1266 consecutive 0.8 mm thick axial slices of the right leg. Each slice had a matrix and in-plane voxel size of 512×512 and 0.43 mm. MRI scan of the subject's right knee was taken with the subject in the supine position and a slightly flexed knee (later computed to be 25.2°) to generate a volumetric dataset for segmentation of the articular and connective tissues. The scan collected 208 sagittal slices that were 0.7 mm thick each and had a matrix size of 512×512 and an in-plane voxel size of 0.31 mm. The MRI scan of the right knee was repeated for three different flexion angles (later computed to be 55.6°, 99.6°, and 137.5°) with the aim of generating a dataset for model validation. The MRI scanner collected 320 consecutive 0.5 mm thick sagittal slices while the matrix and in-plane voxel sizes were 480×480 and 0.5 mm for each slice. During these three acquisitions, the subject was lying in the left lateral recumbent position, the ankle was in a neutral position, and the hip was not flexed to avoid any influence of poly-articular muscles that could affect the patellar position. Finally, to generate a volumetric dataset for segmentation of the muscles in a reasonable amount of time, an MRI scan was used to capture 6.99 mm thick axial slices at pelvic, thigh, knee, and shank levels. Each slice had a matrix size of 512×512 and an in-plane voxel size of 0.97 mm.

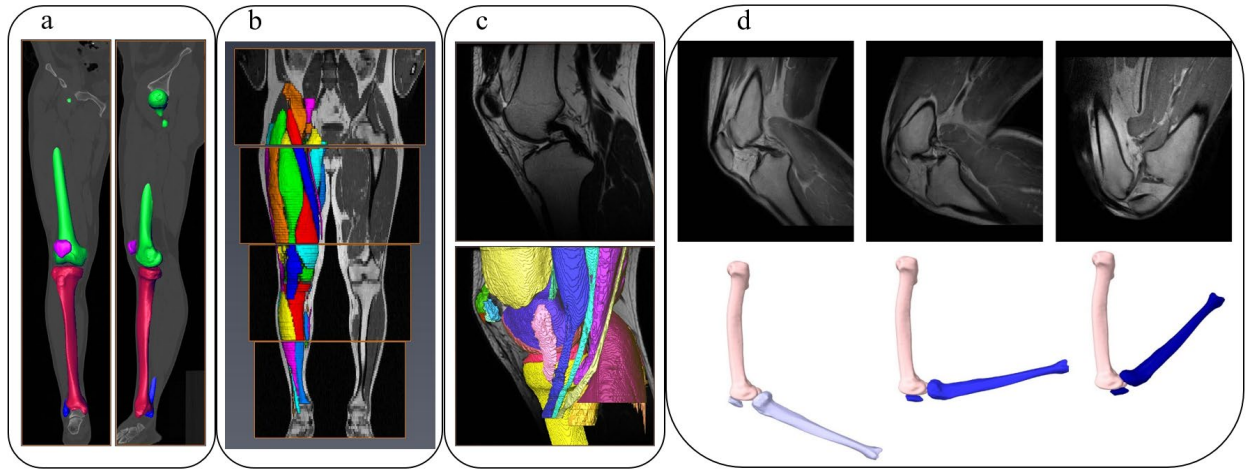

*Figure 1, a) Frontal and sagittal view of the right leg CT scan, b) MRI of the lower limbs and segmented muscle and bones tissues, c) High resolution MRI of the right knee and the segmented tissues demonstrated on the bottom, d) The knee MRI at three different flexion angles and the segmented bones on the bottom*

### 2.2. Geometry reconstruction

Manual segmentation was performed using Amira 6.5.0. The CT scan was used to segment the full bone surfaces in the extended position. These bony surfaces were used to define the bony landmarks required for defining the joint coordinate system and registering the 3D bone shapes segmented from different MRI sets. The MRI scan of the knee at 25.2° flexion and the MRI scan of the full lower limb were segmented to reconstruct the muscles, articular tissues (femoral, patellar, and tibial cartilages), ligaments (patellar ligament and patellar retinacula), and the quadriceps tendon. The MRI scans of the knee that were taken for model validation purposes were used to segment the bones at various knee flexion angles. The segmentation of the patella at each knee flexion angle was repeated three times (twice by a clinician and once by an experienced operator) to account for the impact that segmentation errors can have on the patellofemoral kinematic data points that are used for validating the model. The patella segmented from the CT scan was rigidly registered to the segmented surfaces using Geomagic® Studio 2013 to preserve the exact geometry of the patella between different positions. For this purpose, we used a manual 3-point registration process followed by an automatic global registration, and the standard deviation between the registered surfaces was less than 0.2 mm.

**2.3. Model generation**

A combined FE and multibody model was generated in the Artisynth open-source modeling platform (www.artisynth.org) (Lloyd et al., 2012). The Quadriceps Tendon (QT), Rectus Femoris (RF), Vastus Intermedius (VI), Vastus Medialis (VM), and Vastus Lateralis (VL) muscles, the patellar and femoral cartilages were modeled as 3D FE structures. The bones were modeled as rigid bodies. The medial and lateral patellar retinacula were modeled as they can affect the patellofemoral joint kinematics and tension magnitudes in the patellar ligament (Powers et al., 2006). A discrete number of strands represented the patellar ligament, medial, and lateral patellar retinacula.

Mesh generation was done in HyperMesh 2019 (Altair Engineering, Inc., USA) and produced hexahedral dominant meshes (with optimized element quality and a limited number of wedge elements) for the quadriceps muscles and the quadriceps tendon. The muscular geometries were simplified near their origin and insertion to avoid very small elements. The maximum element size was approximately 6mm in the muscle meshes, and 3mm in the QT mesh. The generated model is illustrated in Figure 2.

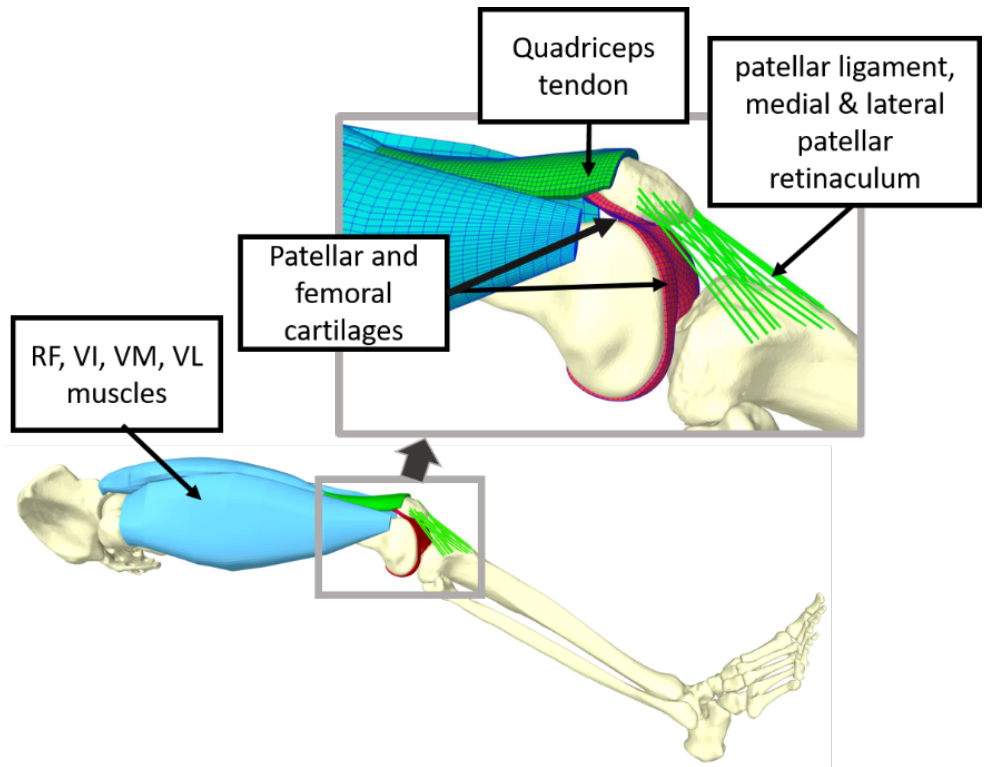

*Figure 2, Generated combined FE-multibody model of the lower limb validated through a simulation of passive deep knee flexion and used for investigating the effect of OWHTO on the patellofemoral joint. The FE model of the quadriceps muscle group included 2100 elements for RF, 1326 for VI, 9290 for VL and 2308 for VM. The QT was formed by 5200 elements, the femoral cartilage had a total of 1106 elements (a single layer of elements) and the patellar cartilage included 969 elements (two layers of elements).*

### 2.4. Material modeling

The constitutive law used to describe the muscle as a continuum material is separated into a passive and an active part. The muscles normally experience large displacements and strains with a non-linear stress/strain curve. Therefore, when the muscle is in the passive state and has no activation, its behavior can be best represented using a hyperelastic material model (Johansson et al., 2000). This passive part can be treated as isotropic or transversely isotropic based on the level of complexity required for the model (Blemker et al., 2005).

Given the objectives of the current study, the muscles were modeled in a passive state, and we assumed that their behavior could be modelled with a nearly incompressible isotropic neo-Hookean hyperelastic material. The parameters used for the quadriceps muscles ($c_1 = 11.7\ kPa$ and $D_1 = 16.3\ MPa^{-1}$, respectively related to the shear modulus and the volumetric variations) were defined based on the findings of *Affagard et al.*, who presented an *in vivo* method to identify the muscle behavior based on a displacement field obtained from ultrasound and digital image correlation techniques (Affagard et al., 2015). For the QT, a linear elastic material was used with a value of 30 MPa for the Young's modulus and 0.46 for the Poisson ratio that was chosen for the model after testing multiple values estimated based on the literature (Weiss and Gardiner, 2001). For the femoral and patellar cartilages, a linear elastic material was used with a Young's modulus of 15MPa and a Poisson ratio of 0.45, similar to what was previously proposed for a study on HTO in the literature (Martay et al., 2018).

Each ligament was represented by a number of strands that had a non-linear stiffening spring behavior at low strains ($\varepsilon$ <0.06) and had a linear stiffness at higher strains (Blankevoort and Huiskes, 1991). In cases where more than one strand was used to model a ligament, the stiffness was divided between the number of strands. The patellar ligament was represented with seven strands (three strands for the central region and two strands for each of the medial and lateral regions). The total stiffness was set to $278\ N/mm$ for the central region and $201\ N/mm$ and $173\ N/mm$ for the medial and lateral regions (Yanke et al., 2015). Five strands represented the medial and lateral patellar retinacula with the respective total stiffness of $31\ N/mm$ and $97\ N/mm$ (LaPrade et al., 2018; Merican et al., 2009). The nominal reference strain was assumed to be 0.02 for all the bundles of the patellar ligament at full extension and 0.0 for the medial and lateral patellar retinacula. The reference length of the bundles (the length at which the springs experience the reference strain) was set to be their length at full extension in the native knee. A 0.02 pre-strain was also imposed along the axial direction of the QT FE component.

### 2.5. Boundary conditions

Sliding contacts were defined between all muscle parts, between the patellar cartilage and the femoral cartilage, between QT and the femoral cartilage, and between the femur and the VM and VL muscles. The nodes of the proximal end of the quadriceps muscle group were attached to the femur or the iliac bone based on the definition of their origins. The nodes of the VI muscle that were closer than one millimeter to the femur surface were attached to the femur. A distance limit was set to attach the nodes on the distal insertion of the muscles to the QT. The nodes located at the distal end of the QT were attached to the patella.

The patella was free to move in all its six degrees of freedom. Gravity was neglected, and the segments' inertial parameters were not considered since the femur and pelvic were non-dynamic.

### 2.6. Simulation of passive deep knee flexion

A simulation of passive deep knee flexion was designed for validation purposes. For that, the position of the tibiofemoral joint at the most flexed position was used to calculate a transformation matrix and control the knee motion. The output of the simulation was the patellofemoral joint kinematics which was then compared against the kinematic data points acquired from the subject's MRI at various knee flexion angles. A schematic view of the setup used for the simulation of passive deep knee flexion is depicted in Figure 3.

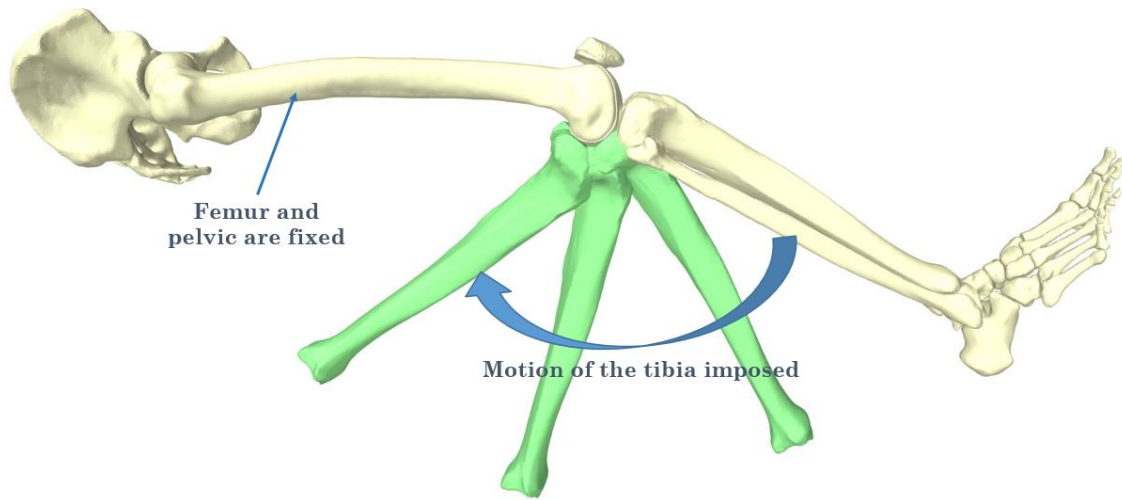

*Figure 3, Schematic view of the passive deep knee simulation setup. The position of the combined shank and foot segment is controlled through the transformation matrix obtained from MRI at different knee flexion angles. The patella is free to move in response to forces transmitted to it through the soft tissues and contacts with the femur. The bone positions at 25.2° knee flexion (the first MRI position) are used to define the initial position of the model.*

### 2.7. Simulation of OWHTO

#### 2.7.1. Virtual OWHTO surgery

Computer models with opened wedges on the tibia were developed for the same subject at 25.2° knee flexion. An oblique cut was simulated on the tibia using a plane perpendicular to the tibial frontal plane and divided the bone into proximal and distal parts. The cut passed through the lateral and medial cortex, respectively 16 mm and 37 mm distal to the tibial plateaus, as inspired from the literature (Akamatsu et al., 2017; Kim et al., 2009).

Models with various angles of wedge opening ranging from 2.5°, 5.0°, 7.5°, to 10.0° were developed using a custom-made Java code in the Artisynth platform as depicted in Figure 4. For simplification, the fibula was attached to the distal part of the tibia. Once the wedge was opened, the distal and proximal parts of the tibia were reunited. To make sure that the wedge opening does alter the limb alignment in other planes than the frontal, the posterior tibial slope (PTS) was checked to remain constant in all cases (approximately 9° that is in the normal range of PTS (10° ± 3.5°) (Paley, 2002)).

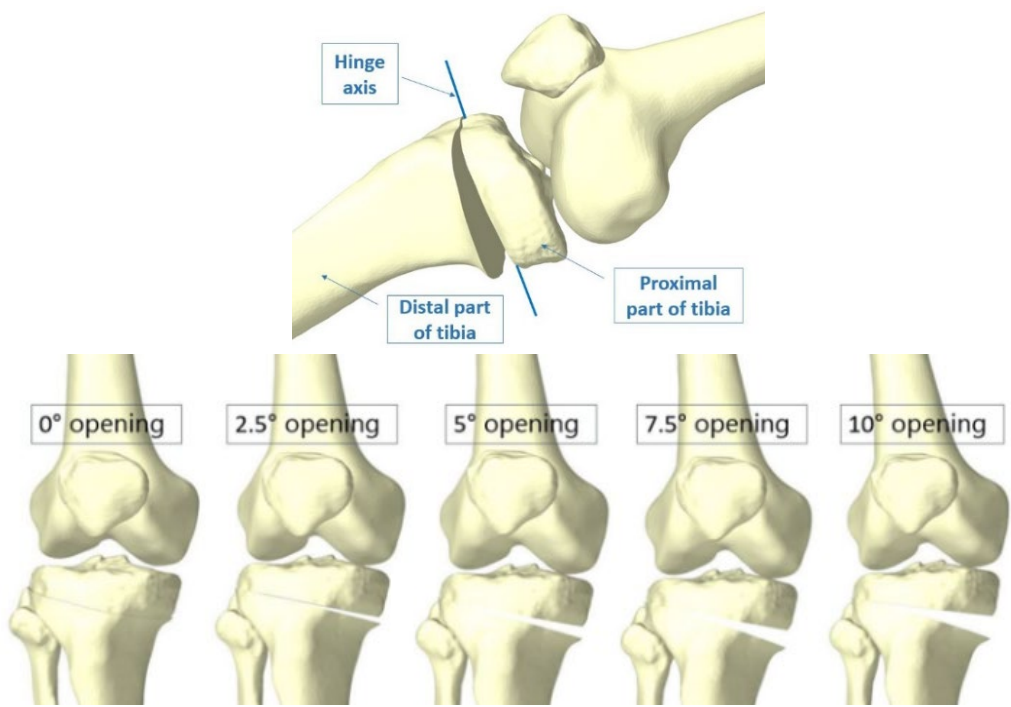

*Figure 4, Top: the virtual wedge opening on the model through cutting the tibia into a proximal and a distal part and rotating the distal part around the defined hinge axis. The hinge axis was defined to be perpendicular to the frontal plane and passing through the most lateral point on the OWHTO cut. Bottom: close view of the tibia after different wedge openings.*

### 2.7.2. OWHTO simulation setup

To perfectly isolate the effect of the distalization of the patellar ligament insertion on the patellar position and the muscles attached to the patella, the knee flexion angle was fixed at 25.2°. The patella was free to react to the forces transmitted to it and find its new equilibrium position after wedge opening. The patellar ligament insertion on the tibia was distal to the osteotomy cut.

Therefore, wedge opening during virtual surgery affected the length and tension in the non-linear springs representing its bundles.

### 2.8. Measurement of the quantities of interest

The bony landmarks and the knee joint coordinate system were defined based on the ISB standard (Grood and Suntay, 1983), as depicted in Figure 5. The origin was placed at the most distal point of the patella to calculate the translations. The kinematics of the patella in all its six degrees of freedom were extracted during model validation with the simulation of passive deep knee flexion. In the post OWHTO models, the impact of wedge opening angle on the patellar position was assessed by monitoring the patellar shift, proximal/distal translation, tilt, and rotation.

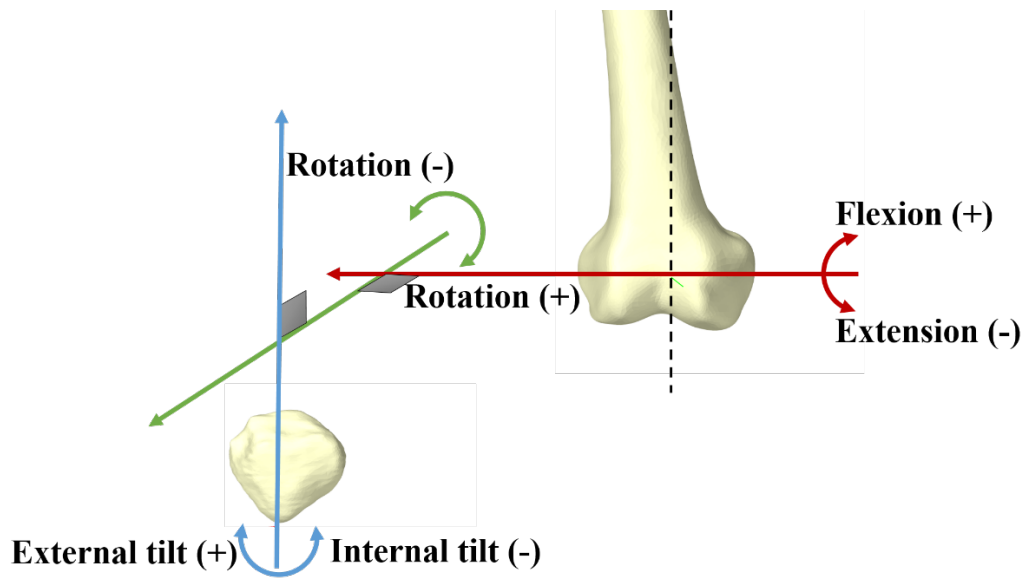

*Figure 5, Description of the patellofemoral JCS: lateral translation or shift is the motion of the patella along the fixed femoral axis (red), anterior translation is along the floating axis (green) and the proximal translation is along the patellar fixed axes (blue).*

## 3. RESULTS

### 3.1. Passive deep knee flexion

A schematic of the comparison between the patellar positions predicted by simulation and taken from MRI segmentation is depicted at the maximal knee flexion position in Figure 6. The patellofemoral kinematic data points acquired from MRI segmentation at various knee flexion angles were consistent with the simulation predictions (Figure 7). However, one exception was the distal patellar translation, which was predicted higher than the MRI values at higher degrees of flexion.

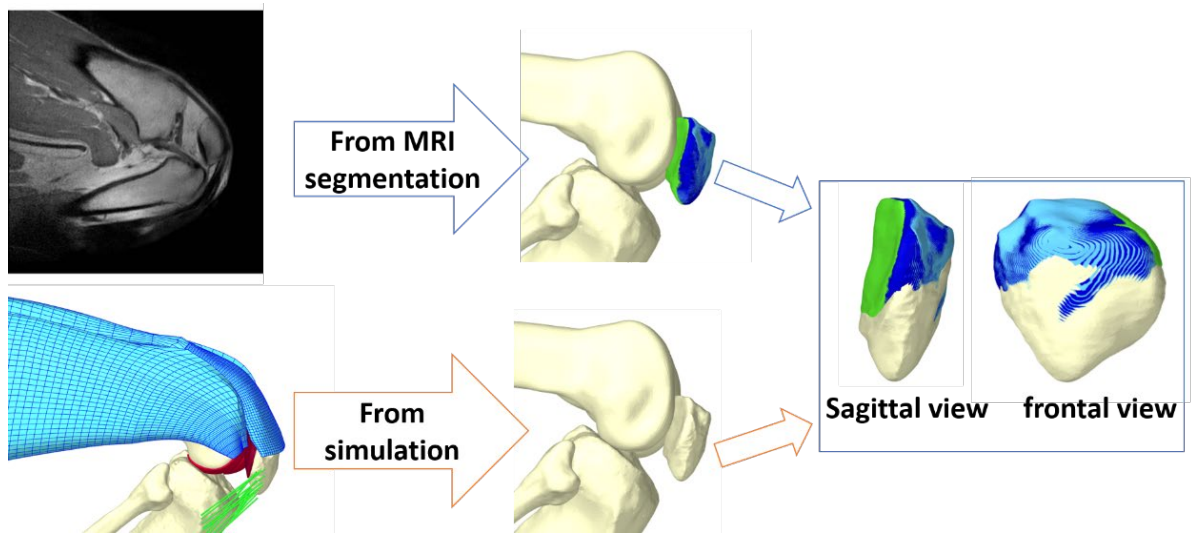

*Figure 6, Schematic view of the patellar position predicted by simulation in comparison with its position acquired from MRI segmentation for 137.5° knee flexion angle. The patella bones depicted in light and dark blue are segmented by the first operator. The patella depicted in green is segmented by the second operator. The patella depicted in white is predicted by simulation.*

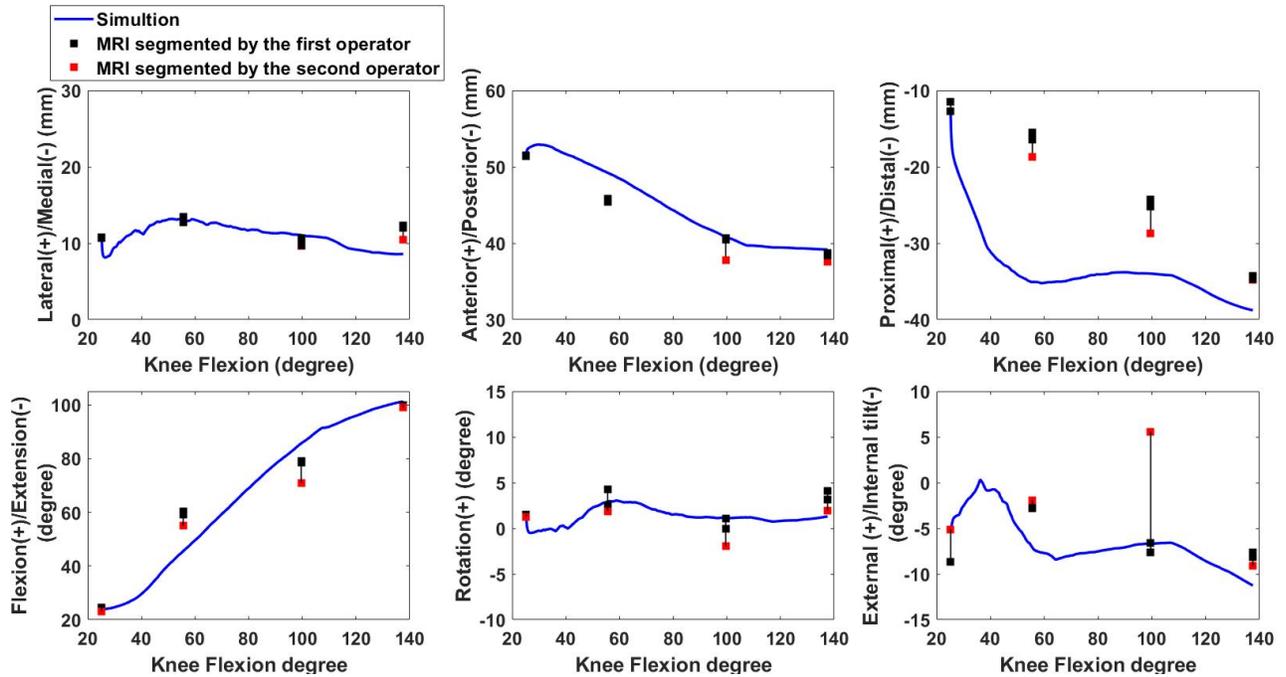

*Figure 7, The patellofemoral kinematics predicted by the simulation of passive deep knee flexion are plotted in blue. The horizontal axis is the knee flexion angle. The vertical axes are the motions of the patella with respect to the femur in its six DOFs. The black and red squares represent the patella kinematics obtained from MRI segmentation respectively done by the first and second operator. The first operator has repeated the segmentation twice to also account for the inter-operator errors. The vertical black lines represent the range of variation between the kinematic points obtained from different MRI segmentations.*

### 3.2. Patellar position after OWHTO

The simulation results showed that the position of the patella in all the analyzed translational and rotational degrees of freedom were affected by the wedge opening angle (translation and rotation alterations respectively reached 5mm and 4°) (Figure 8). The larger wedge openings resulted in distalization of the patella, shifting the patella laterally, rotating the patella, and internal (medial) tilting.

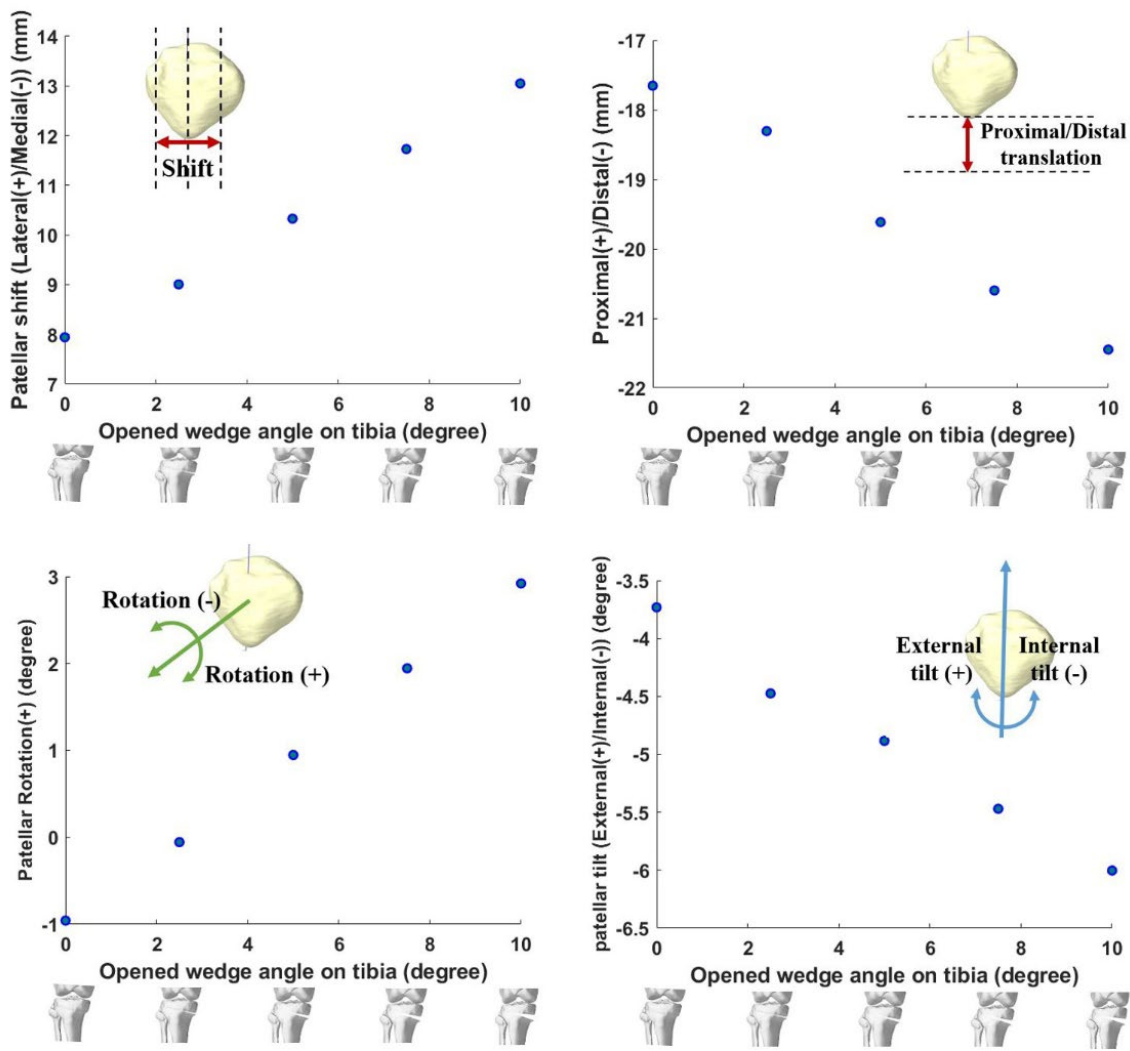

*Figure 8, Predicted patellar position in response to different wedge opening angles on the tibia. The tibiofemoral joint is fixed at 25.2° knee flexion. The horizontal axis shows the wedge opening degree while the vertical axis is the patellar translational and rotational position defined based on ISB standards* (Grood and Suntay 1983).

## 4. DISCUSSION

We have introduced and validated a combined FE-multibody model suitable for simulating the behavior of the patellofemoral joint in various conditions. A key feature of the model is that it can predict the patellofemoral kinematics that result from the interaction of muscles and ligaments attached to the patella and the joint contact forces. We evaluated the model predictions with a simulation of passive deep knee flexion and showed that it produced kinematics consistent with the measures obtained from independent MRI datasets of the same subject at various knee flexion angles.

The validated model was used for predicting patellar position alteration due to different wedge opening angles during an OWHTO surgery at a certain knee flexion angle. The results of the study demonstrated that the OWHTO has a significant effect on the patellofemoral joint. This means that distalization of the patellar ligament insertion due to wedge opening can modify the balance of forces on the patella and alter its position. Our observations also showed that larger wedge openings result in increased values of patellar distalization, lateral patellar shift, patellar rotation, and patellar internal (medial) tilt.

Although our methods for evaluating the change in the height of the patella after OWHTO differ from clinical studies, the qualitative conclusions of our study were consistent with the literature (J. C. Fan, 2012; Longino et al., 2013; Park et al., 2017b). Our findings show that at 25.2° knee flexion, our subject could experience a 3.8 mm post-operative decrease in proximal translation for a 10° wedge opening on the tibia (21.5% decrease from the pre-operative value). While a direct comparison is not possible, this finding is consistent with an MRI-based study (d'Entremont et al., 2014) that found a 14.2% decrease at 30° of flexion among their subject group with various correction angles.

Regarding the alteration of patellar tilt, our results show a maximum internal (or medial) tilt of 2.3° occurring for a 10° wedge opening on the tibia. This finding is consistent with the alterations found in the literature reporting significant decreases of lateral patellar tilt between 1.8° and 2.2° (Bito et al., 2010a; Nha et al., 2016; Park et al., 2017a) or a significant increase of medial patellar tilt of 2.2° with an MRI based method (d'Entremont et al., 2014).

Concerning the patellar shift, our results are consistent with the MRI-based study of 'Entremont et al., 2014 who reported a lateral shift after wedge opening. However, the magnitude of the lateral shift was larger in our case. This is while several clinical papers that used conventional radiology and its corresponding methods were not able to find significant changes in patellar shift (Bito et al., 2010b; Lee et al., 2016; Park et al., 2017b). This ability of the model to correctly predict the lateral shift after wedge opening is of high importance because observations are showing that lateral patellar shift can be associated with patellofemoral pain (MacIntyre et al., 2006).

Our study has the advantage of analyzing the kinematics of the patella in various degrees of freedom while overcoming the limitations of the traditionally used investigation methods. To our knowledge, our study is the first to assess the correlation between wedge angle and patellar position quantitatively. Our analysis underlines the need for reporting the opening angle in future clinical studies rather than only reporting the post-operative standing hip–knee–ankle (HKA) or Weight-bearing Axis (WBA). The correlation between wedge angle and the exact patellar position could be used in the future to explain why performing deformity corrections exceeding a certain wedge opening angle can lead to progression of patellofemoral cartilage injuries (Otakara et al., 2019; Tanaka et al., 2019).

The validated model developed in this study could be used in the future to assess the patellar kinematics after OWHTO over the full range of motion of the knee. Additionally, the stress distribution on the articular and connecting tissues could also be monitored to investigate their relationship with pain and cartilage injury due to the post-operative altered patellar kinematics. Meanwhile, we must acknowledge some of the inherent limitations of this study. The first limitation concerns the simplifications made in the geometry and material model of the simulated muscles. The second limitation concerns the absence of the other stabilization structures connected to the patella, the aponeurosis, the articular capsule of the knee, the skin, and other muscles of the thigh.

## 5. CONCLUSION

We have developed a validated combined FE-multibody model capable of predicting the patellofemoral kinematics. In this model, the patella was in contact with the trochlear groove and was fully free to move in response to the forces transmitted to it via the 3D FE models of the quadriceps muscle group and the ligaments. This detailed model could successfully simulate passive deep knee flexion, and its predictions in terms of patellofemoral kinematics were validated using MR images of the same subject at various knee flexion angles.

The developed model was used for studying the impact of OWHTO on the patellar position at a fixed knee flexion angle and showed that wedge opening could alter the position of the patella. Our observations suggest that larger wedge openings result in increased values of patellar distalization, lateral patellar shift, patellar rotation, and patellar internal (medial) tilt. The contribution of our study is twofold: 1) our results show a correlation between the patellar position and the opening angle on the tibia in various degrees of freedom, and 2) we describe a numerical investigation method based on an explicative biomechanical model which could lead to a better understanding of the effects at play and, eventually, to better clinical outcomes for the patients.

## 6. ACKNOWLEDGMENTS

This work was funded by *Fondation pour la Recherche Médicale* under the project FRM DIC20161236448. The IRMaGe MRI facility was funded by the ANR-11-INSB-0006 grant. The authors wish to acknowledge the continued technical support received from John Lloyd (UBC, Vancouver) throughout this study.

# REFERENCES

Affagard, J.S., Feissel, P., Bensamoun, S.F., 2015. Identification of hyperelastic properties of passive thigh muscle under compression with an inverse method from a displacement field measurement. Journal of Biomechanics 48, 4081–4086. https://doi.org/10.1016/j.jbiomech.2015.10.007

Agarwala, S., Sobti, A., Naik, S., Chaudhari, S., 2016. Comparison of closing-wedge and opening-wedge high tibial osteotomies for medial compartment osteoarthritis of knee in Asian population: Mid-term follow-up. Journal of Clinical Orthopaedics and Trauma 7, 272–275. https://doi.org/10.1016/j.jcot.2016.06.012

Aglietti, P., Buzzi, R., Vena, L.M., Baldini, A., Mondaini, A., 2003. High tibial valgus osteotomy for medial gonarthrosis: a 10- to 21-year study. The journal of knee surgery 16, 21–26.

Akamatsu, Y., Ohno, S., Kobayashi, H., Kusayama, Y., Kumagai, K., Saito, T., 2017. Coronal subluxation of the proximal tibia relative to the distal femur after opening wedge high tibial osteotomy. Knee 24, 70–75. https://doi.org/10.1016/j.knee.2016.09.009

Ali, A.A., Mannen, E.M., Rullkoetter, P.J., Shelburne, K.B., 2020. Validated computational framework for evaluation of in vivo knee mechanics. Journal of Biomechanical Engineering 142. https://doi.org/10.1115/1.4045906

Ali, A.A., Shalhoub, S.S., Cyr, A.J., Fitzpatrick, C.K., Maletsky, L.P., Rullkoetter, P.J., Shelburne, K.B., 2016. Validation of predicted patellofemoral mechanics in a finite element model of the healthy and cruciate-deficient knee. Journal of Biomechanics 49, 302–309. https://doi.org/10.1016/j.jbiomech.2015.12.020

Bito, H., Takeuchi, R., Kumagai, K., Aratake, M., Saito, I., Hayashi, R., Sasaki, Y., Saito, T., 2010a. Opening wedge high tibial osteotomy affects both the lateral patellar tilt and patellar height. Knee Surgery, Sports Traumatology, Arthroscopy 18, 955–960. https://doi.org/10.1007/s00167-010-1077-5

Bito, H., Takeuchi, R., Kumagai, K., Aratake, M., Saito, I., Hayashi, R., Sasaki, Y., Saito, T., 2010b. Opening wedge high tibial osteotomy affects both the lateral patellar tilt and patellar height. Knee Surgery, Sports Traumatology, Arthroscopy 18, 955–960. https://doi.org/10.1007/s00167-010-1077-5

Blankevoort, L., Huiskes, R., 1991. Ligament-bone interaction in a three-dimensional model of the knee. Journal of Biomechanical Engineering 113, 263–269. https://doi.org/10.1115/1.2894883

Blemker, S.S., Pinsky, P.M., Delp, S.L., 2005. A 3D model of muscle reveals the causes of nonuniform strains in the biceps brachii. Journal of Biomechanics 38, 657–665. https://doi.org/10.1016/j.jbiomech.2004.04.009

Coventry, M.B., Ilstrup, D.M., Wallrichs, S.L., 1993. Proximal tibial osteotomy: A critical long-term study of eighty-seven cases. Journal of Bone and Joint Surgery - Series A 75, 196–201. https://doi.org/10.2106/00004623-199302000-00006

d'Entremont, A.G., McCormack, R.G., Horlick, S.G.D., Stone, T.B., Manzary, M.M., Wilson, D.R., 2014. Effect of opening-wedge high tibial osteotomy on the three-dimensional kinematics of the knee. The Bone & Joint Journal 96-B, 1214–1221. https://doi.org/10.1302/0301-620X.96B9.32522

Duivenvoorden, T., Brouwer, R.W., Baan, A., Bos, P.K., Reijman, M., Bierma-Zeinstra, S.M.A., Verhaar, J.A.N., 2014. Comparison of closing-wedge and opening-wedge high tibial osteotomy for medial compartment osteoarthritis of the knee: A randomized controlled trial with a six-year follow-up. Journal of Bone and Joint Surgery - American Volume 96, 1425–1432. https://doi.org/10.2106/JBJS.M.00786

Elyasi, E., Cavalié, G., Perrier, A., Graff, W., Payan, Y., 2021. A Systematic Review on Selected Complications of Open-Wedge High Tibial Osteotomy from Clinical and Biomechanical Perspectives. Applied Bionics and Biomechanics 2021, 1–14. https://doi.org/10.1155/2021/9974666

Erdemir, A., Guess, T.M., Halloran, J., Tadepalli, S.C., Morrison, T.M., 2012. Considerations for reporting finite element analysis studies in biomechanics. https://doi.org/10.1016/j.jbiomech.2011.11.038

Fan, J.C., 2012. Open wedge high tibial osteotomy: cause of patellar descent. Journal of Orthopaedic Surgery and Research 7, 3. https://doi.org/10.1186/1749-799X-7-3

Fan, J.C.H., 2012. Open wedge high tibial osteotomy: Cause of patellar descent. Journal of Orthopaedic Surgery and Research 7. https://doi.org/10.1186/1749-799X-7-3

Goshima, K., Sawaguchi, T., Shigemoto, K., Iwai, S., Nakanishi, A., Ueoka, K., 2017. Patellofemoral Osteoarthritis Progression and Alignment Changes after Open-Wedge High Tibial Osteotomy Do Not Affect Clinical Outcomes at Mid-term Follow-up. Arthroscopy - Journal of Arthroscopic and Related Surgery 33, 1832–1839. https://doi.org/10.1016/j.arthro.2017.04.007

Grood, E.S., Suntay, W.J., 1983. A joint coordinate system for the clinical description of three-dimensional motions: Application to the knee. Journal of Biomechanical Engineering 105, 136–144. https://doi.org/10.1115/1.3138397

Hernigou, P., Medevielle, D., Debeyre, J., Goutallier, D., 1987. Proximal tibial osteotomy for osteoarthritis with varus deformity. A ten to thirteen-year follow-up study. Journal of Bone and Joint Surgery - Series A 69, 332–354. https://doi.org/10.2106/00004623-198769030-00005

Javidan, P., Adamson, G.J., Miller, J.R., Durand, P., Dawson, P.A., Pink, M.M., Lee, T.Q., 2013. The effect of medial opening wedge proximal tibial osteotomy on patellofemoral contact. American Journal of Sports Medicine 41, 80–86. https://doi.org/10.1177/0363546512462810

Johansson, T., Meier, P., Blickhan, R., 2000. A Finite-Element Model for the mechanical analysis of Skeletal Muscles. J Theor Biol. https://doi.org/10.1006/jtbi.2000.2109\rS0022-5193(00)92109-X [pii]

Kim, K. Il, Kim, D.K., Song, S.J., Lee, S.H., Bae, D.K., 2017. Medial Open-Wedge High Tibial Osteotomy May Adversely Affect the Patellofemoral Joint. Arthroscopy - Journal of Arthroscopic and Related Surgery 33, 811–816. https://doi.org/10.1016/j.arthro.2016.09.034

Kim, S.J., Koh, Y.G., Chun, Y.M., Kim, Y.C., Park, Y.S., Sung, C.H., 2009. Medial opening wedge high-tibial osteotomy using a kinematic navigation system versus a conventional method: A 1-year retrospective, comparative study. Knee Surgery, Sports Traumatology, Arthroscopy 17, 128–134. https://doi.org/10.1007/s00167-008-0630-y

LaPrade, M.D., Kallenbach, S.L., Aman, Z.S., Moatshe, G., Storaci, H.W., Turnbull, T.L., Arendt, E.A., Chahla, J., LaPrade, R.F., 2018. Biomechanical Evaluation of the Medial Stabilizers of the Patella. American Journal of Sports Medicine 46, 1575–1582. https://doi.org/10.1177/0363546518758654

Lee, Y.S., Lee, S.B., Oh, W.S., Kwon, Y.E., Lee, B.K., 2016. Changes in patellofemoral alignment do not cause clinical impact after open-wedge high tibial osteotomy. Knee Surgery, Sports Traumatology, Arthroscopy 24, 129–133. https://doi.org/10.1007/s00167-014-3349-y

Lloyd, J.E., Stavness, I., Fels, S., 2012. ArtiSynth: A Fast Interactive Biomechanical Modeling Toolkit Combining Multibody and Finite Element Simulation, in: Studies in Mechanobiology, Tissue Engineering and Biomaterials. Springer, pp. 355–394. https://doi.org/10.1007/8415_2012_126

Loia, M.C., Vanni, S., Rosso, F., Bonasia, D.E., Bruzzone, M., Dettoni, F., Rossi, R., 2016. High tibial osteotomy in varus knees: Indications and limits. Joints 4, 98–110. https://doi.org/10.11138/jts/2016.4.2.098

Longino, P.D., Birmingham, T.B., Schultz, W.J., Moyer, R.F., Giffin, J.R., 2013. Combined tibial tubercle osteotomy with medial opening wedge high tibial osteotomy minimizes changes in patellar height: A prospective cohort study with historical controls. American Journal of Sports Medicine 41, 2849–2857. https://doi.org/10.1177/0363546513505077

MacIntyre, N.J., Hill, N.A., Fellows, R.A., Ellis, R.E., Wilson, D.R., 2006. Patellofemoral joint kinematics in individuals with and without patellofemoral pain syndrome. Journal of Bone and Joint Surgery - Series A 88, 2596–2605. https://doi.org/10.2106/JBJS.E.00674

Martay, J.L., Palmer, A.J., Bangerter, N.K., Clare, S., Monk, A.P., Brown, C.P., Price, A.J., 2018. A preliminary modeling investigation into the safe correction zone for high tibial osteotomy. Knee 25, 286–295. https://doi.org/10.1016/j.knee.2017.12.006

Merican, A.M., Sanghavi, S., Iranpour, F., Amis, A.A., 2009. The structural properties of the lateral retinaculum and capsular complex of the knee. Journal of Biomechanics 42, 2323–2329. https://doi.org/10.1016/j.jbiomech.2009.06.049

Müller, J.H., Razu, S., Erdemir, A., Guess, T.M., 2020. Prediction of patellofemoral joint kinematics and contact through co-simulation of rigid body dynamics and non-linear finite element analysis. Computer Methods in Biomechanics and Biomedical Engineering 23, 718–733. https://doi.org/10.1080/10255842.2020.1761960

Nha, K.-W., Kim, H.-J., Ahn, H.-S., Lee, D.-H., 2016. Change in Posterior Tibial Slope After Open-Wedge and Closed-Wedge High Tibial Osteotomy. The American Journal of Sports Medicine 44, 3006–3013. https://doi.org/10.1177/0363546515626172

Noyes, F.R., Mayfield, W., Barber-Westin, S.D., Albright, J.C., Heckmann, T.P., 2006. Opening wedge high tibial osteotomy: An operative technique and rehabilitation program to decrease complications and promote early union and function. American Journal of Sports Medicine 34, 1262–1273. https://doi.org/10.1177/0363546505286144

Otakara, E., Nakagawa, S., Arai, Y., Inoue, H., Kan, H., Nakayama, Y., Fujii, Y., Ueshima, K., Ikoma, K., Fujiwara, H., Kubo, T., 2019. Large deformity correction in medial open-wedge high tibial osteotomy may cause degeneration of patellofemoral cartilage: A retrospective study. Medicine 98, e14299. https://doi.org/10.1097/MD.0000000000014299

Paley, D., 2002. Principles of Deformity Correction, Principles of Deformity Correction. Springer Berlin Heidelberg. https://doi.org/10.1007/978-3-642-59373-4

Park, H., Kim, H.W., Kam, J.H., Lee, D.H., 2017a. Open Wedge High Tibial Osteotomy with Distal Tubercle Osteotomy Lessens Change in Patellar Position. BioMed Research International 2017. https://doi.org/10.1155/2017/4636809

Park, H., Kim, H.W., Kam, J.H., Lee, D.H., 2017b. Open Wedge High Tibial Osteotomy with Distal Tubercle Osteotomy Lessens Change in Patellar Position. BioMed Research International 2017. https://doi.org/10.1155/2017/4636809

Powers, C.M., Chen, Y.J., Farrokhi, S., Lee, T.Q., 2006. Role of peripatellar retinaculum in transmission of forces within the extensor mechanism. Journal of Bone and Joint Surgery - Series A 88, 2042–2048. https://doi.org/10.2106/JBJS.E.00929

Song, I.H., Song, E.K., Seo, H.Y., Lee, K.B., Yim, J.H., Seon, J.K., 2012. Patellofemoral alignment and anterior knee pain after closing- and opening-wedge valgus high tibial osteotomy. Arthroscopy - Journal of Arthroscopic and Related Surgery 28, 1087–1093. https://doi.org/10.1016/j.arthro.2012.02.002

Stoffel, K., Willers, C., Korshid, O., Kuster, M., 2007. Patellofemoral contact pressure following high tibial osteotomy: A cadaveric study. Knee Surgery, Sports Traumatology, Arthroscopy 15, 1094–1100. https://doi.org/10.1007/s00167-007-0297-9

Sun, H., Zhou, L., Li, F., Duan, J., 2017. Comparison between Closing-Wedge and Opening-Wedge High Tibial Osteotomy in Patients with Medial Knee Osteoarthritis: A Systematic Review and Meta-analysis. Journal of Knee Surgery 30, 158–165. https://doi.org/10.1055/s-0036-1584189

Tanaka, T., Matsushita, T., Miyaji, N., Ibaraki, K., Nishida, K., Oka, S., Araki, D., Kanzaki, N., Hoshino, Y., Matsumoto, T., Kuroda, R., 2019. Deterioration of patellofemoral cartilage status after medial open-wedge high tibial osteotomy. Knee Surgery, Sports Traumatology, Arthroscopy 27, 1347–1354. https://doi.org/10.1007/s00167-018-5128-7

Weiss, J.A., Gardiner, J.C., 2001. Computational modeling of ligament mechanics. Critical Reviews in Biomedical Engineering. https://doi.org/10.1615/CritRevBiomedEng.v29.i3.20

Yanke, A., Bell, R., Lee, A., Shewman, E.F., Wang, V., Bach, B.R., 2015. Regional mechanical properties of human patellar tendon allografts. Knee Surgery, Sports Traumatology, Arthroscopy 23, 961–967. https://doi.org/10.1007/s00167-013-2768-5